\documentclass[jmp,%
 aip,
 amsmath,amssymb,
 reprint,%
]{revtex4-1}

\usepackage{graphicx}
\usepackage{dcolumn}
\usepackage{bm}
\usepackage{hyperref}
\usepackage{moresize}
\usepackage{amsmath}
\usepackage[utf8, applemac]{inputenc}
\usepackage[T1]{fontenc}
\usepackage{mathptmx}
\usepackage{etoolbox}
\usepackage{amssymb}
\usepackage{xcolor}
\newcommand{\mycirc}[1][black]{{\textcolor{#1}{\ensuremath\bullet}}}
\definecolor{1purple}{RGB}{85,63,150}
\definecolor{1blue}{RGB}{8,57,148}
\definecolor{1pink}{RGB}{249,211,189}
\definecolor{1wheat}{RGB}{196,196,27}
\definecolor{1grey}{RGB}{187,187,187}
\definecolor{1red}{RGB}{231,20,26}
\definecolor{1green}{RGB}{33,146,75}
\definecolor{1cyan}{RGB}{111,199,213}
\newcommand{\tabref}[1]{\mbox{Table~\ref{#1}}}

\renewcommand{\eqref}[1]{\mbox{Eq.~(\ref{#1})}} 
\newcommand{\figpanel}[2]{Fig.~\hyperref[#1]{\ref*{#1}(#2)}} 
\newcommand{\figpanels}[3]{Fig.~\hyperref[#1]{\ref*{#1}(#2)-(#3)}} 
\newcommand{\figpanelNoPrefix}[2]{\hyperref[#1]{\ref*{#1}(#2)}} 
\newcommand{\figpanelsNoPrefix}[3]{\hyperref[#1]{\ref*{#1}(#2)-(#3)}} 

\usepackage{mleftright} 
\newcommand{\abs}[1]{\mleft|#1\mright|}
\newcommand{\abssq}[1]{\mleft| #1 \mright|^2}

\usepackage{units}
\hypersetup{%
	plainpages=true,
	breaklinks=true,
	hypertexnames=false,
	pageanchor=true,
	colorlinks=true,
	linkcolor={blue},
	citecolor={blue},
	urlcolor={blue},
	anchorcolor={black}
}
\usepackage[all]{hypcap} 

\makeatletter
\def\@email#1#2{%
 \endgroup
 \patchcmd{\titleblock@produce}
  {\frontmatter@RRAPformat}
  {\frontmatter@RRAPformat{\produce@RRAP{*#1\href{mailto:#2}{#2}}}\frontmatter@RRAPformat}
  {}{}
}%
\makeatother
\begin{document}

\preprint{AIP/123-QED}

\title{Tunable coherent microwave beam splitter and combiner at the single-photon level} 



\author{Y.-H.~Huang}
\affiliation{Department of Physics, National Tsing Hua University, Hsinchu 30013, Taiwan}
\author{K.-M.~Hsieh}
\affiliation{Department of Physics, City University of Hong Kong, Kowloon, Hong Kong SAR 999077, China}%
\author{F.~Aziz}
\affiliation{Department of Physics, National Tsing Hua University, Hsinchu 30013, Taiwan}
\author{Z.~Q.~Niu}
\affiliation{State Key Laboratory of Materials for Integrated Circuits, Shanghai Institute of Microsystem and Information Technology, Chinese Academy of Sciences, Shanghai 200050, China}
\affiliation{ShanghaiTech University, Shanghai 201210, China}
\author{P.~Y.~Wen}
\affiliation{Department of Physics, National Chung Cheng University, Chiayi 621301, Taiwan}
\author{Y.-T.~Cheng}
\affiliation{Department of Physics, City University of Hong Kong, Kowloon, Hong Kong SAR 999077, China}
\author{Y.-S.~Tsai}
\affiliation{Department of Physics, National Tsing Hua University, Hsinchu 30013, Taiwan}
\author{J.~C.~Chen}
\affiliation{Department of Physics, National Tsing Hua University, Hsinchu 30013, Taiwan}
\affiliation{Center for Quantum Technology, National Tsing Hua University, Hsinchu 30013, Taiwan}
\author{Xin Wang}
\affiliation{Department of Physics, City University of Hong Kong, Kowloon, Hong Kong SAR 999077, China}
\author{A.~F.~Kockum}
\affiliation{Department of Microtechnology and Nanoscience, Chalmers University of Technology, 412 96 Gothenburg, Sweden}
\author{Z.-R.~Lin}
\thanks{Authors to whom correspondence should be addressed: \\
zrlin@mail.sim.ac.cn; \\
yhlin@phys.nthu.edu.tw; \\
iochoi@cityu.edu.hk}
\affiliation{State Key Laboratory of Materials for Integrated Circuits, Shanghai Institute of Microsystem and Information Technology, Chinese Academy of Sciences, Shanghai 200050, China}
\author{Y.-H.~Lin}
\thanks{Authors to whom correspondence should be addressed: \\
zrlin@mail.sim.ac.cn; \\
yhlin@phys.nthu.edu.tw; \\
iochoi@cityu.edu.hk}
\affiliation{Department of Physics, National Tsing Hua University, Hsinchu 30013, Taiwan}
\affiliation{Center for Quantum Technology, National Tsing Hua University, Hsinchu 30013, Taiwan}
\author{I.-C.~Hoi}
\thanks{Authors to whom correspondence should be addressed: \\
zrlin@mail.sim.ac.cn; \\
yhlin@phys.nthu.edu.tw; \\
iochoi@cityu.edu.hk}
\affiliation{Department of Physics, National Tsing Hua University, Hsinchu 30013, Taiwan}
\affiliation{Department of Physics, City University of Hong Kong, Kowloon, Hong Kong SAR 999077, China}

\date{\today}


\begin{abstract}
A beam splitter is a key component used to direct and combine light paths in various optical and microwave systems. It plays a crucial role in devices like interferometers, such as the Mach--Zehnder and Hong--Ou--Mandel setups, where it splits light into different paths for interference measurements. These measurements are vital for precise phase and coherence testing in both classical and quantum optical experiments. In this work, we present a nonlinear beam splitter and beam combiner utilizing a frequency-tunable superconducting artificial atom in a one-dimensional open waveguide. This beam splitter is highly versatile, with adjustable transparency ranging from unity to zero for signals at the single-photon level. Additionally, the beam combiner can merge two coherent beams, generating interference fringes as the relative phase between them varies.
\end{abstract}

\pacs{}

\maketitle 




Beam splitters and combiners are fundamental optical and microwave components that manipulate light by either splitting an incoming beam into two paths or merging them into a single one. In linear optical quantum computing (LOQC)~\cite{Knill2001, Braunstein2005, Kok2007, rohde2012optical, paesani2021scheme, madsen2022quantum, graham2022multi}, beam splitters and combiners are crucial building blocks for implementing quantum logic gates using photons; any unitary $N \times N$ matrix can be implemented with an array of fewer than $N^2/2$ beam-splitters and some phase shifters~\cite{Reck1994, Bouland2014}. Beam combiners are essential for observing the Hong--Ou--Mandel effect~\cite{hong1987measurement}, which plays a significant role in verifying photon indistinguishability. In interferometry, beam splitters divide light beams into separate paths, allowing them to interfere upon recombination and enabling high-precision measurements of distances~\cite{lee2013absolute, yang2015absolute}, surface profiles~\cite{farahi2018inverse}, and even gravitational waves~\cite{abbott2009ligo, abbott2016observation}. A classic example of their application in precision metrology is the Mach--Zehnder interferometer~\cite{zhang2024unbalanced} (MZI), which relies on beam splitters and beam combiners to achieve interference-based measurements.

In the search for significant light-matter interactions that can aid LOQC, superconducting circuits~\cite{Gu2017, blais2021circuit}, often referred to as artificial atoms, offer a distinctive and versatile platform for implementing beam splitters and combiners at microwave frequencies. A significant advantage of these circuits, which have been considered for LOQC~\cite{Adhikari2013}, is the ability to engineer and customize the parameters of artificial atoms and the light-matter interaction. Recent efforts have focused on achieving strong coupling between superconducting artificial atoms and propagating microwave photons, leading to the emergence of waveguide quantum electrodynamics (wQED)~\cite{Roy2017, Sheremet2023} with superconducting qubits. This field has seen remarkable advances, including resonance fluorescence~\cite{astafiev2010resonance}, photon routing~\cite{hoi2011demonstration}, non-classical microwaves~\cite{hoi2012generation}, Landau--Zener--St\"uckelberg--Majorana interferometry~\cite{Wen2020}, cross-Kerr effect between two fields~\cite{hoi2013giant}, lifetime control of artificial atoms~\cite{hoi2015probing}, atom-field interaction control via phase shaping~\cite{Cheng2024}, amplification without population inversion~\cite{wen2018reflective}, collective Lamb shift between two superconducting qubits~\cite{wen2019large}, coherent dynamics of a photon-dressed qubit~\cite{Liul2023}, giant atoms~\cite{kannan2020waveguide}, photon-mediated interactions between atoms~\cite{van2013photon}, deterministic photon loading~\cite{lin2022deterministic}, etc.

These pioneering experiments have inspired new directions in demonstrating microwave components for interferometers within the framework of wQED using superconducting circuits. In this framework, beam splitting has been achieved~\cite{Gu2017} with, e.g., two transmission lines close to each other~\cite{gabelli2004hanbury}, a 90- or 180-degree hybrid coupler~\cite{Mariantoni2005, Pozar2011}, a Wilkinson power divider~\cite{Mariantoni2010, Pozar2011}, or a three-wave mixer with Josephson junctions~\cite{Abdo2013}. However, these approaches tend to suffer from requiring complicated setups or lacking tunability.

In this Letter, we demonstrate a transmon qubit~\cite{koch2007charge}, an artificial atom, functioning as a simple tunable beam splitter and combiner. A few theoretical works have considered using such an artificial atom~\cite{Roulet2016}, or an artificial atom coupled to a cavity~\cite{Oehri2015}, as a beam splitter for single microwave photons. In Ref.~\cite{hoi2011demonstration}, Hoi \textit{et al}.~demonstrated experimentally the transmittance and reflectance of a transmon as a function of control pulse power, accompanied by incoherent loss that led to a very lossy beam splitter. Here, we engineer a transmon qubit to function as a highly efficient, negligible-loss beam splitter with tunable transparency controlled by adjusting an external magnetic flux. Additionally, we engineer the transmon qubit to combine two incoming coherent light beams, producing an interference pattern that varies with the relative phase between the two incoming beams. These findings demonstrate the potential of superconducting circuits for driving the development of microwave-based MZIs and advancing quantum computing with linear optical systems.

\begin{figure}
\includegraphics[width=\linewidth]{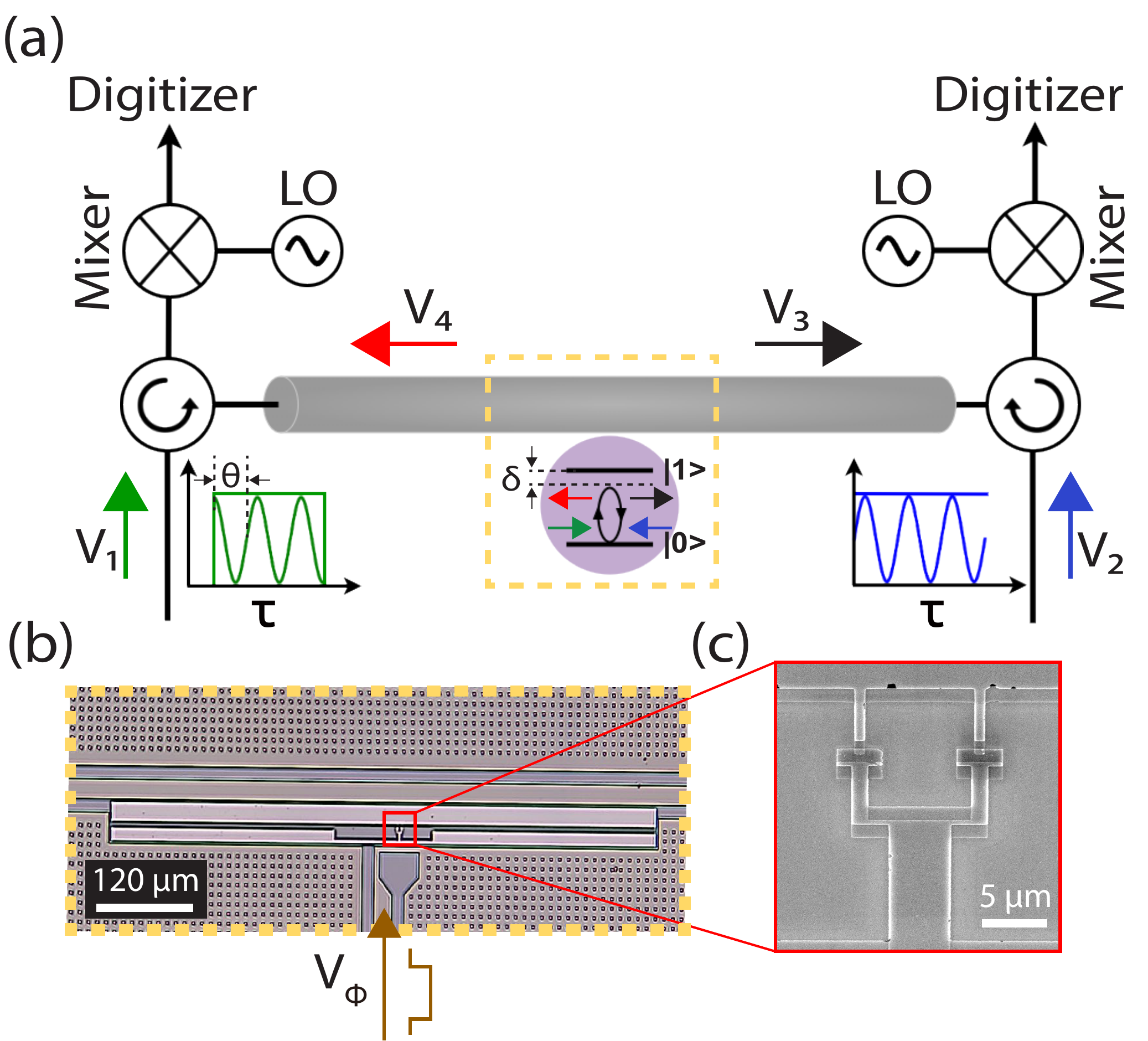}
\caption{The experimental scheme and device.
(a) A simplified schematic of the experiment (see Fig.~S1 in the Supplementary Material for details). An artificial atom (purple disk inside the yellow dashed square) is embedded in a 1D open transmission line. Two input waves, $V_1$ (green arrow) and $V_2$ (blue arrow), with a carrier frequency detuned by $\delta$ from the resonance frequency of the atom, can be simultaneously applied to the atom. The output signals $V_3$ (black arrow) and $V_4$ (red arrow) are down-converted by a local oscillator (LO) and recorded by a digitizer.
(b) Optical micrograph of the device. A transmon qubit (light gray) is capacitively coupled to a 1D open transmission line (dark gray). The red square indicates the SQUID.
(c) Scanning electron micrograph of the SQUID.
\label{Fig1}}
\end{figure}

\begin{table}
    \centering
    \begin{tabular}{|c|c|c|c|c|}
    \hline
        $\omega/2\pi$ & $\Gamma/2\pi$ & $\Gamma^n/2\pi$ & $\gamma/2\pi$ & $\phi$ \\ \hline
        [GHz] & [MHz] & [MHz] & [MHz] & $[rad]$ \\ \hline
        $4.1108$ & $22.15$ & $0.39$ & $11.47$ & $0.0526$ \\ \hline
    \end{tabular}
\caption{Extracted qubit parameters from single-tone spectroscopy. The detailed experimental setup is described in Fig.~S1 of the Supplementary Material.}
\label{tab:2}
\end{table}

\figpanel{Fig1}{a} illustrates the configuration for our on-chip beam splitter and combiner. The device incorporates a transmon qubit with a superconducting quantum interference device (SQUID) loop, capacitively coupled to the center of a one-dimensional (1D) open transmission line. We first characterize our device using transmission spectroscopy to extract the essential parameters at the qubit's resonance frequency $\omega / 2\pi = \unit[4.1108]{GHz}$ (see \tabref{tab:2}) for time-domain measurements and simulation. Here, $\Gamma$ and $\gamma = \Gamma / 2 + \Gamma^n$ are the relaxation and decoherence rates, respectively; the non-radiative decoherence rate $\Gamma^n$ (including pure dephasing and intrinsic loss) is much smaller than $\Gamma$. The detailed characterization method is given in Sec.~S2 of the Supplementary Material.

For the beam-splitter operation, a single input field [denoted $V_1$, indicated by the green arrow in \figpanel{Fig1}{a}] is introduced. A cryogenic microwave circulator directs this field and facilitates its interaction with the two-level system (atom), as indicated by the yellow dashed box in \figpanel{Fig1}{a}. We then measure the resulting transmitted ($V_3$, black arrow) and reflected ($V_4$, red arrow) output fields via heterodyne detection. In the beam combiner operation, two input fields, $V_1$ and $V_2$, are introduced with a phase $\theta$ in $V_1$ that can be varied. The combined fields are again the outgoing fields $V_3$ and $V_4$.

The mathematical representation of the reciprocal beam splitter and beam combiner is a $2 \times 2$ matrix~\cite{gerry2023introductory} relating the output fields ($V_3$, $V_4$) to the input fields ($V_1$, $V_2$) as a function of the phase $\theta$ and the time $\tau$:
\begin{align}
\begin{pmatrix}
V_3 \\
V_4
\end{pmatrix}
&=
\begin{pmatrix}
t_L & r_R \\
r_L & t_R 
\end{pmatrix}
\begin{pmatrix}
V_1 \\
V_2
\end{pmatrix} 
\label{eq:IOmatrix} , \\
V_1(\theta,\tau) &= V_{1, \text{off}} \: e^{i (\omega_p \tau + \theta + \theta_0)} , \label{eq:V1V2} \\
V_2(\tau) &= V_{2, \text{off}} \: e^{i \omega_p \tau} . 
\end{align}
Here, $V_{1, \text{off}}$ and $V_{2, \text{off}}$ represent the bare on-chip amplitudes of the incoming waves without any interaction with the qubit and $\theta_0$ is the initial phase difference between $V_1$ and $V_2$. The calibration of these parameters is described in Sec.~S3 in the Supplementary Material.

The transmission coefficient $t$ and reflection coefficient $r$\cite{lu2021characterizing} are given by
\begin{align}
\label{eq:tr}
t &= 1 + r , \\
\label{eq:r}
r &= -e^{i \phi} \frac{\Gamma}{2 \gamma} \frac{1 - i \frac{\delta}{\gamma}}{1 + \mleft( \frac{\delta}{\gamma} \mright)^2 + \frac{\Omega^2}{\gamma \Gamma}} .
\end{align}
Here, $\delta = \omega_p - \omega$, is the detuning between the qubit frequency $\omega$ and the probe frequency $\omega_p$, $\Omega$ is the Rabi frequency, and $\phi$ is the impedance mismatch angle. The subscripts $R$ and $L$ in \eqref{eq:IOmatrix} refer to inputs on the right side and the left side of the qubit, respectively. Ideally, when both the transmittance $\abssq{t} = T$ and the reflectance $\abssq{r} = R$ are $0.5$, maximum or minimum visibility will be observed depending on whether the interference is constructive or destructive as the relative phase $\theta$ changes.

\begin{figure*}
\includegraphics[width=\linewidth]{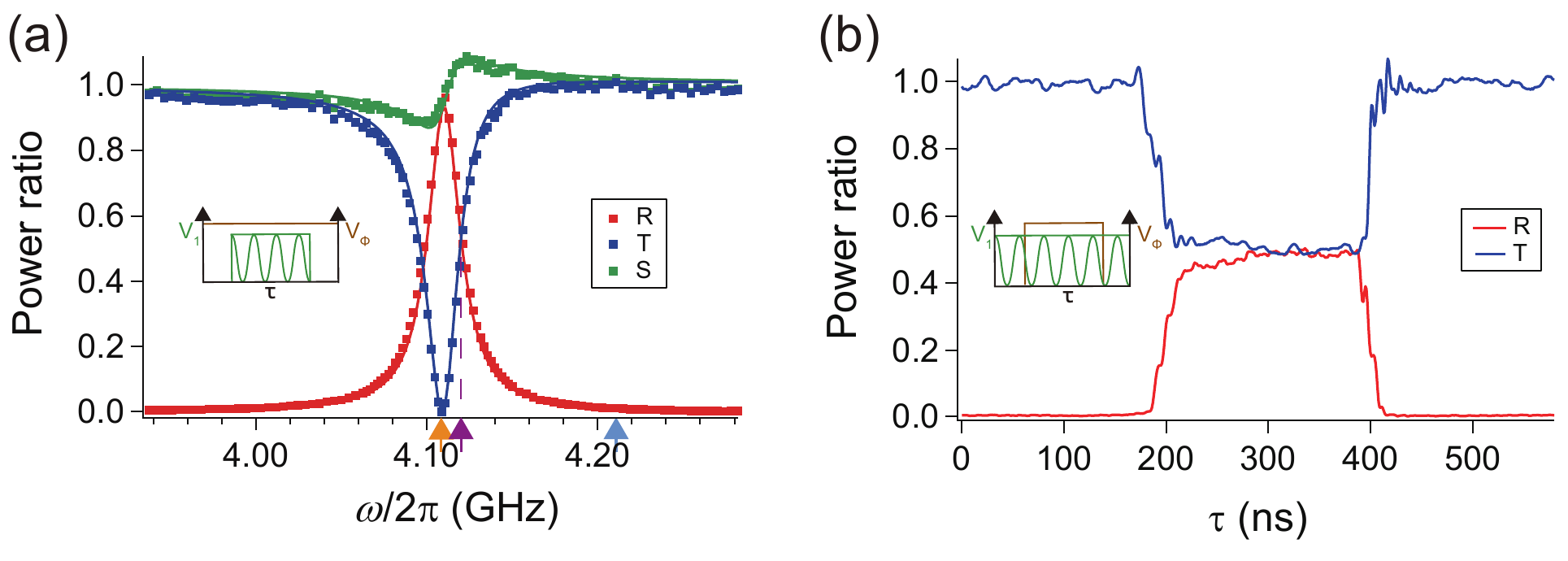}
\caption{A tunable beam splitter.
(a) We apply a weak pulse as input (green wave in the inset) and record the transmittance $T$ and reflectance $R$. We set $\omega_p / 2\pi = \unit[4.1108]{GHz}$ (orange arrow) and vary $\omega / 2\pi$ by applying a global flux $V_\Phi$ (brown wave in the inset) to manipulate the transparency of the qubit.
The red (blue) dots are the reflectance (transmittance) data, which are the averages of the steady-state region across the pulse at different $\omega / 2\pi$. The green dots $S$ are the sum of the reflectance and the transmittance. The solid curves represent the theoretical predictions, according to Eqs.~(\ref{eq:tr}) and (\ref{eq:r}), with each curve matching a corresponding color.
(b) Transmittance and reflectance using a continuous-wave input (green wave in the inset) with $\omega_p / 2\pi = \unit[4.1108]{GHz}$ [orange arrow in \figpanel{Fig2}{a}]. A Gaussian square pulse (brown wave in the inset) serves as a flux pulse to rapidly tune the qubit from $\omega / 2\pi = \unit[4.2108]{GHz}$ [cyan arrow in \figpanel{Fig2}{a}] to $\omega / 2\pi = \unit[4.12]{GHz}$ [purple arrow in \figpanel{Fig2}{a}]. With these settings, we observe a ratio of reflectance and transmittance around 50:50 during the square pulse.
\label{Fig2}}
\end{figure*}

We now explore the qubit's functionality as a beam splitter using a time-domain setup [see \figpanel{Fig1}{a}]. A pulse [green in the inset of \figpanel{Fig2}{a}] with $\omega_p / 2\pi = \unit[4.1108]{GHz}$ [orange arrow in \figpanel{Fig2}{a}] is applied to the qubit. In this experiment, we control the resonance frequency $\omega$ of the artificial atom by applying an external flux $V_\Phi$ [see \figpanel{Fig1}{b}]. As seen in \figpanel{Fig2}{a}, the artificial atom exhibits near $\unit[100]{\%}$ ($\unit[0]{\%}$) reflectance (transmittance) when the input is exactly resonant ($\delta = 0$). Conversely, when the atom is far detuned, it becomes off-resonant ($\abs{\delta} \gg \gamma$) with the incoming wave, resulting in no interaction, where $T\sim\unit[100]{\%}$ and $R\sim\unit[0]{\%}$. Nearly resonant tuning ($\abs{\delta}\sim\gamma$) within the qubit's linewidth results in partial transmittance and partial reflectance.

In \figpanel{Fig2}{a}, the blue and red dots are experimental data averaged over the steady-state region in the output pulses, while the solid curves are theoretical fits using Eqs.~(\ref{eq:tr}) and (\ref{eq:r}), showing good agreement. The green dots and curve depict the experimental data and theoretical fit of the sum of transmittance and reflectance ($S = T + R$), respectively. In the vicinity of the resonance frequency, we see $S \neq 1$, which we attribute to impedance mismatch ($\phi \neq 0$) from the transmission line. 

The data in \figpanel{Fig2}{a} show that the qubit functions as a tunable beam splitter, with transparency that can be adjusted from unity to zero by tuning the external flux. Furthermore, applying a flux pulse [brown in the inset of \figpanel{Fig2}{b}] to rapidly tune the qubit from $\omega / 2\pi = \unit[4.2108]{GHz}$ [cyan arrow in \figpanel{Fig2}{a}] to $\omega / 2\pi = \unit[4.12]{GHz}$ [purple arrow in \figpanel{Fig2}{a}] achieves a $50:50$ ratio of transmittance and reflectance, demonstrating the transparency tunability within a few nanoseconds. The flux-pulse amplitude can be adjusted to modify $T$ and $R$ (see Fig.~S4 in the Supplementary Material for details).

\begin{figure*}
\includegraphics[width=\linewidth]{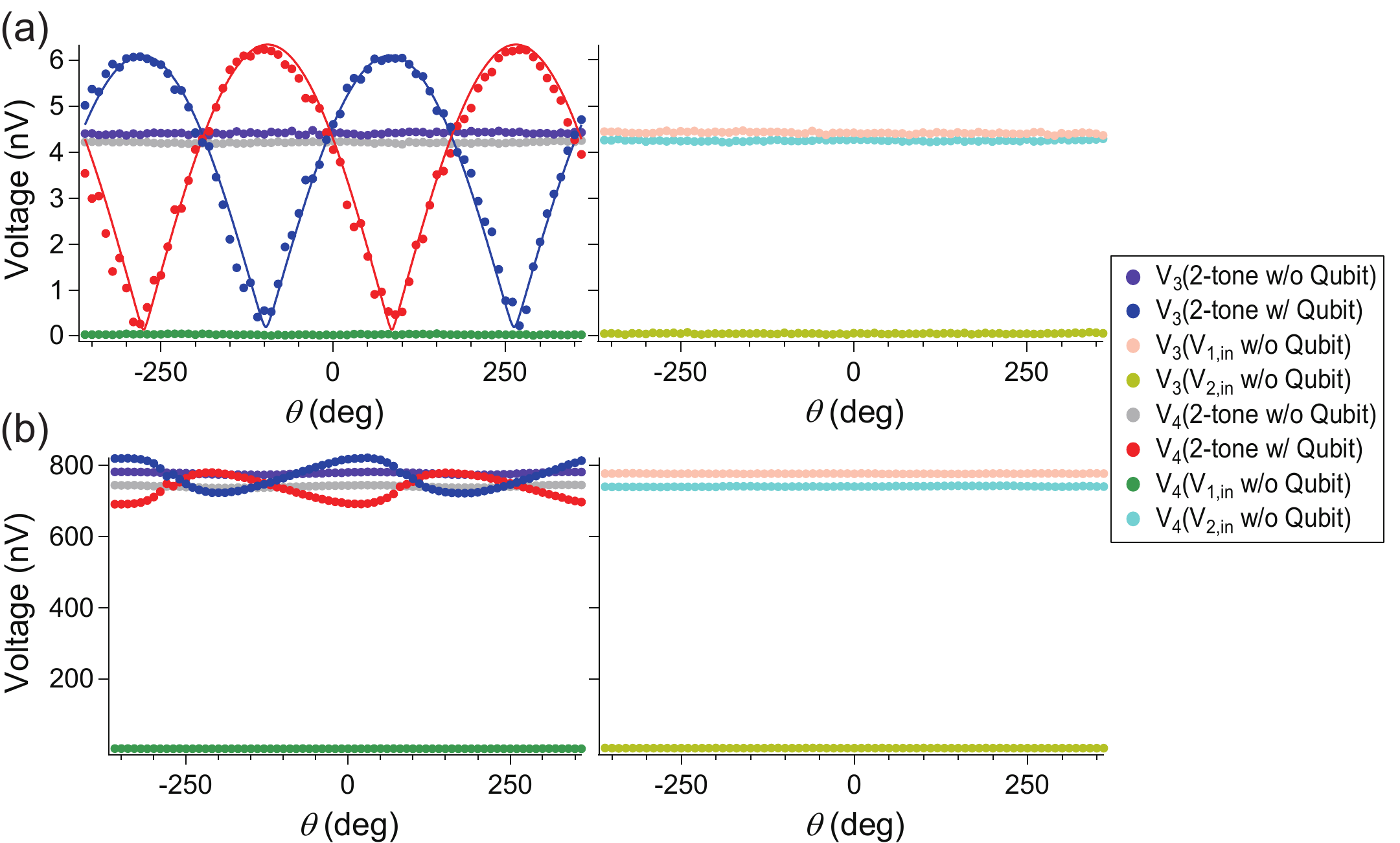}
\caption{A nonlinear tunable beam combiner. 
(a) When the qubit is tuned to $\omega / 2\pi = \unit[4.12]{GHz}$ [purple arrow in \figpanel{Fig2}{a}], the ratio of reflectance and transmittance is around 50:50. Two weak input waves, $V_1$ [green arrow in \figpanel{Fig1}{a}] and $V_2$ [blue arrow in \figpanel{Fig1}{a}] with $\omega_p / 2\pi = \unit[4.1108]{GHz}$, are simultaneously sent to the qubit. The output signals oscillate as a function of the phase $\theta$ of $V_1$ [as defined in \figpanel{Fig1}{a}]. The blue \mycirc[1blue] and red \mycirc[1red] data are the experimental data when the qubit is tuned to $\omega / 2\pi = \unit[4.12]{GHz}$; the solid curves with the corresponding colors are the theory predictions from Eqs.~(\ref{eq:IOmatrix})--(\ref{eq:r}) using the parameters in \tabref{tab:3}. The purple \mycirc[1purple] (gray \mycirc[1grey]) data are the $V_3$ ($V_4$) signals when the qubit is far detuned. The pink \mycirc[1pink] (green \mycirc[1green]) data are the $V_3$ ($V_4$) signals when the qubit is far detuned and there only $V_1$ is applied to the qubit. The wheat \mycirc[1wheat] (cyan \mycirc[1cyan]) data are the $V_3$ ($V_4$) signals when the qubit is far detuned and only $V_2$ is applied to the qubit.
(b) In the same setup as shown in \figpanel{Fig3}{a}, but with increased probe power ($\Omega \gg \gamma$), the interference visibility decreases dramatically. The remaining oscillations indicate that the high probe power does not completely saturate the qubit.
\label{Fig3}}
\end{figure*}

\begin{table*}
    \centering
    \begin{tabular}{|c|c|c|c|c|c|c|}
    \hline
        $V_{\rm 1,off}$ & $V_{\rm 2,off}$ & $t_L$ & $r_R$ & $r_L$ & $t_R$ & $\theta_0$\\ \hline
        [nV] & [nV] & - & - & - & - & [rad] \\ \hline
        $4.261+1.129i$ & $2.182+3.634i$ & $0.674+0.241i$ & $-0.387+0.577i$ & $-0.707+0.203i$ & $-0.010+0.730i$ & $1.152$\\ \hline
    \end{tabular}
\caption{Scattering-matrix elements at $\omega_p / 2\pi = \unit[4.1108]{GHz}$ and $\omega / 2\pi = \unit[4.12]{GHz}$ for \figpanel{Fig3}{a}. See Sec.~S4 in the Supplementary Material for details.
\label{tab:3}}
\end{table*}

Next, we investigate the qubit's functionality as a beam combiner. We set the qubit's resonance frequency to $\omega / 2\pi = \unit[4.12]{GHz}$ [purple arrow in \figpanel{Fig2}{a}], achieving a transmittance to reflectance ratio of approximately $50:50$. Two weak probes $V_1$ and $V_2$ at $\omega_p / 2\pi = \unit[4.1108]{GHz}$ are simultaneously applied to the atom. The output signals oscillate as a function of the phase $\theta$ of $V_1$, as shown in \figpanel{Fig3}{a}. In \figpanel{Fig3}{a}, the blue and red dots represent the experimental data $V_3$ and $V_4$, while the solid curves of corresponding colors are the theoretical predictions based on Eqs.~(\ref{eq:IOmatrix})--(\ref{eq:r}) (see Sec.~S4 in the Supplementary Material for details; the scattering-matrix elements are summarized in \tabref{tab:3}). 

When the qubit is far detuned, as shown by the purple (gray) dots for $V_3$ ($V_4$), the input fields no longer interact with the qubit, resulting in no interference between the two input fields. This is evident as the pink and cyan dots in \figpanel{Fig3}{a}, where these single-tone data remain constant irrespective of changes in $\theta$. Imperfect isolation of the cryogenic microwave circulator causes leakage signals, mitigated using a phase shifter, tunable attenuator, and directional coupler (see Sec.~S1 in the Supplementary Material), with results shown by the wheat and green dots at negligible constant levels in \figpanel{Fig3}{a}. 

Above, we examined the beam combiner with $T \sim 0.5$, where the qubit was subjected to low probe power. Now, we increase the input pulse power  by \unit[45]{dB}, from low to high such that $\Omega \gg \gamma$, while keeping all other parameters unchanged. In this configuration, the field is primarily transmitted, since the artificial atom is almost saturated, and only a weak interference pattern is observed, as shown in \figpanel{Fig3}{b}, with reduced visibility compared to the results in \figpanel{Fig3}{a}. However, the remaining oscillations in the blue ($V_3$) and red ($V_4$) data points indicate that the high probe power does not completely saturate the qubit. To confirm this observation, we conducted an experiment with a far-detuned qubit (i.e., without qubit interaction), which revealed identical power levels in the output signals $V_3$ and $V_4$, with no interference effects present [see purple and gray data in \figpanel{Fig3}{b}].

In conclusion, we have demonstrated that a transmon qubit in a 1D open transmission line can function as a nonlinear beam splitter and combiner at the single-photon level in the microwave regime. By adjusting an external magnetic field, we can tune the qubit's resonance frequency and modify the ratio of transmittance and reflectance of the interacting light. In this way, we can also widely adjust the working frequency range. Additionally, we showed that we can combine two incoming coherent lights, with interference fringes appearing as a function of the relative phase between them. These results highlight the robustness, simplicity, and versatility of a superconducting artificial atom as both a beam splitter and combiner, which are critical components of quantum-technology applications such as MZIs or LOQC.

See the Supplementary Material for details on the schematic of the experimental setup, qubit characterization via the single-tone spectroscopy, calibration measurement of the required parameters, scattering matrix elements calculation, and the detailed data for \figpanel{Fig2}{b} and \figpanel{Fig3}{a}.


\begin{acknowledgments}
\indent I.-C.H.~acknowledges financial support from City University of Hong Kong project 9610617, from the Research Grants Council of Hong Kong (Grant No.~11307324) and from Guangdong Provincial Quantum Science Strategic Initiative (Grant No. GDZX2203001, GDZX2303005, GDZX2403001). AFK acknowledges support from the Swedish Foundation for Strategic Research (grant numbers FFL21-0279 and FUS21-0063), the Horizon Europe programme HORIZON-CL4-2022-QUANTUM-01-SGA via the project 101113946 OpenSuperQPlus100, and from the Knut and Alice Wallenberg Foundation through the Wallenberg Centre for Quantum Technology (WACQT).
\end{acknowledgments}


\section*{Author Declarations}
\vspace{-5mm}
\subsection*{Conflict of Interest} 
\vspace{-5mm}
The authors have no conflicts to disclose.

\subsection*{Author Contributions}
Y.-H.~Huang, K.-M.~Hsieh, F.~Aziz and Z.~Q.~Niu contributed equally to this work.

\textbf{Y.-H.~Huang}: Conceptualization (equal); Data curation (lead); Writing - original draft (equal); Writing - review \& editing (equal); Methodology (equal); Formal analysis (equal).
\textbf{K.-M.~Hsieh}: Conceptualization (equal); Data curation (lead); Writing - original draft (equal); Writing - review \& editing (equal); Methodology (equal); Formal analysis (equal).
\textbf{F.~Aziz}: Conceptualization (equal); Data curation (equal); Writing - original draft (equal); Writing - review \& editing (equal).
\textbf{Z.~Q.~Niu}: Conceptualization (equal); Resources (equal); Writing - review \& editing (equal).
\textbf{P.~Y.~Wen}: Conceptualization (equal). 
\textbf{Y.-T.~Cheng}: Conceptualization (equal); Software (lead). 
\textbf{Y.-S.~Tsai}: Conceptualization (equal). 
\textbf{J.~C.~Chen}: Conceptualization (equal). 
\textbf{Xin Wang}: Writing - review \& editing (equal); Funding acquisition (equal). 
\textbf{A.~F.~Kockum}: Conceptualization (equal); Writing - review \& editing (equal); Funding acquisition (equal). 
\textbf{Z.-R.~Lin}: Conceptualization (equal); Resources (equal); Supervision (equal).
\textbf{Y.-H.~Lin}: Conceptualization (equal); Supervision (equal).
\textbf{I.-C.~Hoi}: Conceptualization (equal); Supervision (lead); Writing - review \& editing (equal); Project administration (lead); Funding acquisition (equal).


\section*{Data Availability}
The data that support the findings of this study are available from the corresponding author upon reasonable request.

\section*{REFERENCES}

\bibliography{main}

\end{document}


\title[]{Supplementary Material for ``Tunable coherent microwave beam splitter and combiner at the single-photon level''}
\author{Y.-H.~Huang}
\homepage{These authors contributed equally to this work.}
\affiliation{Department of Physics, National Tsing Hua University, Hsinchu 30013, Taiwan}
 
 \author{K.-M.~Hsieh}
\homepage{These authors contributed equally to this work.}
\affiliation{Department of Physics, City University of Hong Kong, Kowloon, Hong Kong SAR 999077, China}

\author{F.~Aziz}
\homepage{These authors contributed equally to this work.}
\affiliation{Department of Physics, National Tsing Hua University, Hsinchu 30013, Taiwan}

\author{Z.~Q.~Niu}
\homepage{These authors contributed equally to this work.}
\affiliation{State Key Laboratory of Materials for Integrated Circuits, Shanghai Institute of Microsystem and Information Technology,
Chinese Academy of Sciences, Shanghai 200050, China}
\affiliation{ShanghaiTech University, Shanghai 201210, China}

\author{P.~Y.~Wen}
\affiliation{Department of Physics, National Chung Cheng University, Chiayi 621301, Taiwan}

\author{Y.-T.~Cheng}
\affiliation{Department of Physics, City University of Hong Kong, Kowloon, Hong Kong SAR 999077, China}

\author{Y.-S.~Tsai}
\affiliation{Department of Physics, National Tsing Hua University, Hsinchu 30013, Taiwan}

\author{J.~C.~Chen}
\affiliation{Department of Physics, National Tsing Hua University, Hsinchu 30013, Taiwan}
\affiliation{Center for Quantum Technology, National Tsing Hua University, Hsinchu 30013, Taiwan}

\author{Xin Wang}
\affiliation{Department of Physics, City University of Hong Kong, Kowloon, Hong Kong SAR 999077, China}

\author{A.~F.~Kockum}
\affiliation{Department of Microtechnology and Nanoscience, Chalmers University of Technology, 412 96 Gothenburg, Sweden}

\author{Z.-R.~Lin}
\homepage{Corresponding authors: zrlin@mail.sim.ac.cn, yhlin@phys.nthu.edu.tw, iochoi@cityu.edu.hk}
\affiliation{State Key Laboratory of Materials for Integrated Circuits, Shanghai Institute of Microsystem and Information Technology,
Chinese Academy of Sciences, Shanghai 200050, China}

\author{Y.-H.~Lin}
\homepage{Corresponding authors: zrlin@mail.sim.ac.cn, yhlin@phys.nthu.edu.tw, iochoi@cityu.edu.hk}
\affiliation{Department of Physics, National Tsing Hua University, Hsinchu 30013, Taiwan}
\affiliation{Center for Quantum Technology, National Tsing Hua University, Hsinchu 30013, Taiwan}

\author{I.-C.~Hoi}
\homepage{Corresponding authors: zrlin@mail.sim.ac.cn, yhlin@phys.nthu.edu.tw, iochoi@cityu.edu.hk}
\affiliation{Department of Physics, National Tsing Hua University, Hsinchu 30013, Taiwan}
\affiliation{Department of Physics, City University of Hong Kong, Kowloon, Hong Kong SAR 999077, China}


\maketitle

\tableofcontents

\renewcommand{\thefigure}{S\arabic{figure}}
\renewcommand{\thesection}{S\arabic{section}}
\renewcommand{\theequation}{S\arabic{equation}}
\renewcommand{\thetable}{S\arabic{table}}
\renewcommand{\bibnumfmt}[1]{[S#1]~}
\renewcommand{\citenumfont}[1]{S#1}


\newpage

\section{Details of the experimental setup}
\label{sec:1}

\begin{figure}
\includegraphics[width=\linewidth]{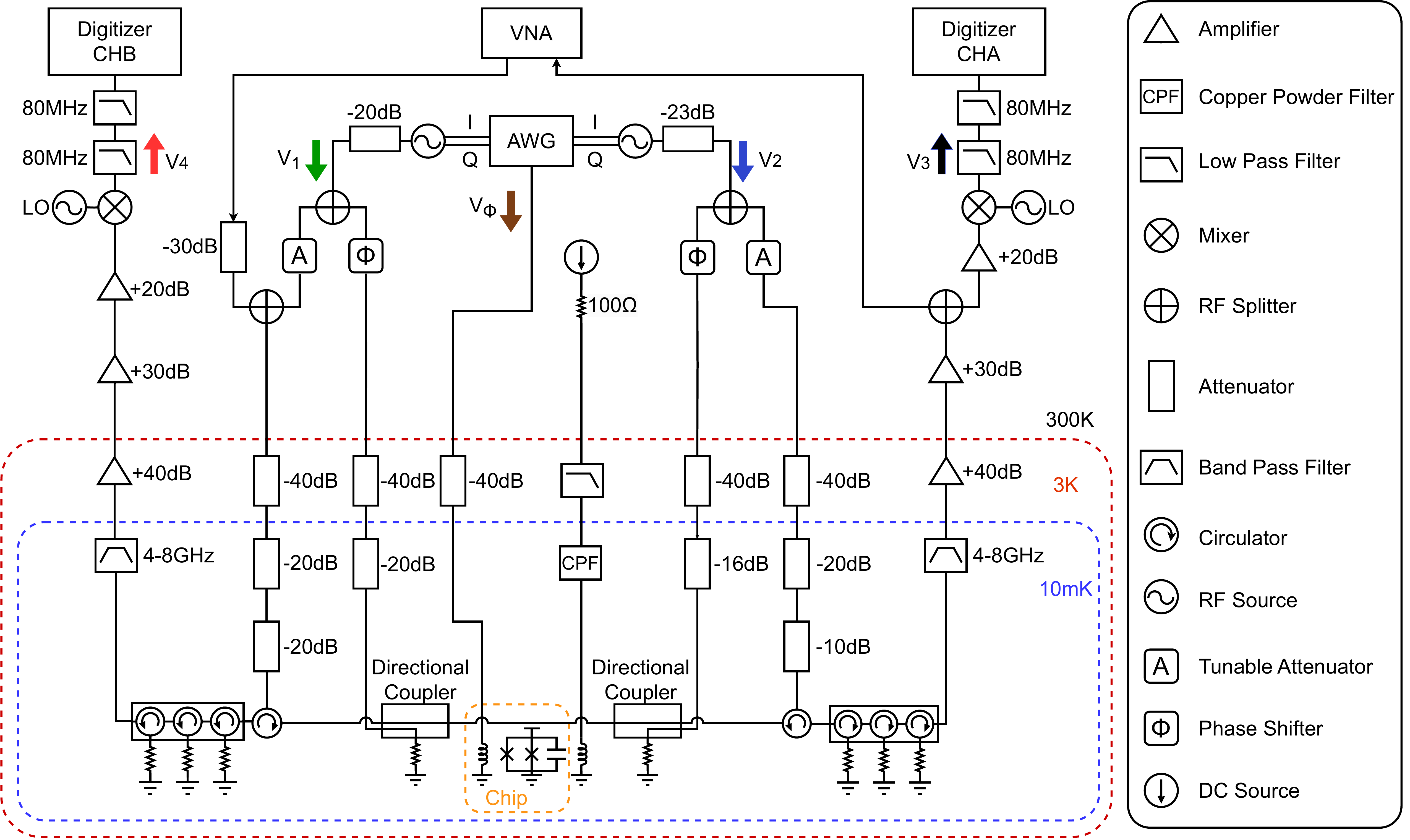}
\caption{Schematic of the experimental setup. The superconducting artificial atom, a transmon qubit, is in the orange dashed box. The transmon is regulated by room-temperature electronics, including a flux line, input lines, and readout lines. The blue and red dashed boxes indicate different temperature stages within the dilution refrigerator.
\label{fig:1}}
\end{figure}

The experimental setup, illustrated in \figref{fig:1}, consists of a sample box containing a chip (highlighted by an orange dashed box) that features a single artificial atom capacitively coupled to an open one-dimensional transmission line. This assembly is mounted at the bottom of a dilution refrigerator operating at a base temperature of \unit[10]{mK}. A superconducting magnetic coil, centrally positioned on the sample box, is used to tune the resonance frequency of the artificial atom. This coil is controlled by a DC source at room temperature, connected through a low-pass filter (LPF) and a copper powder filter (CPF) for noise filtering. The experimental setup is symmetrically arranged on both sides of the sample box to facilitate both frequency-domain and time-domain measurements, supporting our beam-splitter and beam-combiner experiments. 

Single-tone spectroscopy is conducted using a vector network analyzer (VNA) to characterize the qubit, detailed further in \secref{sec:2}. During these experiments, a weak probe field is introduced through the VNA at a probe frequency $\omega_p$, attenuated at several stages within the dilution refrigerator. A cryogenic circulator then directs this input field to interact with the qubit. The field transmitted from the qubit is captured by the VNA after passing through a series of cryogenic circulators, a band-pass filter (BPF), and a two-stage amplification process involving a high-electron-mobility transistor (HEMT) amplifier at \unit[3]{K} and another amplifier at room temperature.

For time-domain measurements, an arbitrary waveform generator (AWG) generates IQ signals for the IQ modulator of the radio-frequency (RF) source to produce shaped modulation signals. The modulation signals are input as a pulse on the left side ($V_1$, green arrow), and/or a continuous wave on the right side ($V_2$, blue arrow). The output signals, $V_3$ (black arrow) and $V_4$ (red arrow), on both sides are measured at digitizer channels $A$ and $B$ (CHA and CHB), respectively. Additionally, an RF splitter is utilized at the input sides of the modulation signals. One of the split fields is routed to the cryogenic circulator through several attenuators and a tunable attenuator, denoted by A, at room temperature. The other field is connected to the directional coupler through several attenuators and a phase shifter, denoted as $\Phi$, at room temperature. This configuration mitigates leakage from the imperfect isolation of the circulators. The detailed procedure for cancellation of leakage fields can be found in Ref.~\cite{Hoi_2013}. The near-perfect cancellation of leakage field can be seen in Fig.~3 [$V_3$ ($V_{2, \text{in}}$ w/o qubit) and $V_{4}$ ($V_{1, \text{in}}$ w/o qubit)] in the main text. The local flux through the chip (inside the orange-dashed box) is controlled by the AWG via the $V_\Phi$ line (brown arrow) for fast switching, with a $\unit[-40]{dB}$ attenuation at \unit[3]{K}.


\section{Single-tone spectroscopy in the frequency domain}
\label{sec:2}

\begin{figure}
\includegraphics[width=\linewidth]{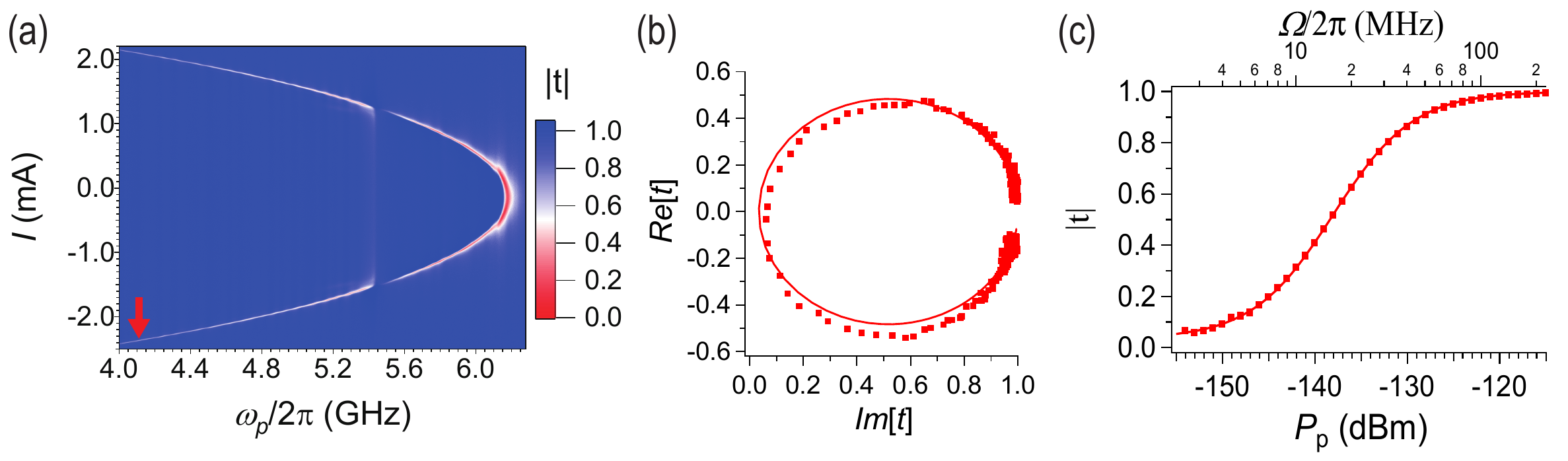}
\caption{Single-tone frequency-domain measurement of the transmission coefficient.
(a) Transmission coefficient as a function of probe frequency $\omega_p$ and the current $I$ controlling the flux through the SQUID of the transmon qubit. 
(b) Transmission coefficient plotted in the IQ plane for a weak probe (i.e., assuming $\Omega \ll \gamma$).
(c) The on-resonance transmission magnitude $\abs{t}$ as a function of probe power $P_p (\propto \Omega^2)$. The solid curves in panels (a) and (b) are theory using Eqs.~(4) and (5) in the main text.
\label{fig:2}}
\end{figure}

To characterize the parameters of the qubit, we conducted single-tone spectroscopy. \figpanel{fig:2}{a} illustrates the magnitude of the transmission coefficient, $\abs{t}$, as a function of the probe frequency $\omega_p$ and the current $I$ applied in the superconducting coil to control the global flux that sets the transition frequencies of the transmon. The spectrum reveals the cosine dependence on the flux of the qubit's $\ket{0} \leftrightarrow \ket{1}$ transition frequency. We note that a stray cavity resonance is visible at $\omega_p / 2 \pi = \unit[5.425]{GHz}$, accompanied by additional box modes appearing as vertical lines near the stray cavity and the apex. To avoid interference from these modes, we selected a biasing point for our beam-splitter and -combiner experiments at $\omega_p / 2 \pi = \unit[4.1108]{GHz}$, marked by a red arrow.

\begin{table*}
\centering
\begin{tabular}{| c | c | c | c | c | c | c  | c | c |}
\hline
$E_C / h$  & $E_J / h$ & $E_J / E_C$ & $\omega/ 2 \pi$  &  $\Gamma / 2 \pi$  & $\Gamma_{}^{n}/2\pi$ & $\gamma / 2 \pi$ & $\phi$ \\
\hline
[MHz] & [GHz]  & - & [GHz] & [MHz]  & [MHz] & [MHz] & [rad] \\
\hline
199.4 & 25.57 & 128.24 & 4.1108 & 22.15  & 0.39 & 11.47 & 0.0526 \\
\hline
\end{tabular}
\caption{Extracted qubit parameters. We obtain the resonance frequency $\omega$, the relaxation rate $\Gamma$, and the decoherence rate $\gamma$ from \figpanel{fig:2}{b} by circle fitting in the IQ plane~\cite{Probst2015, lu2021characterizing}. The non-radiative decay rate $\Gamma^n$ is calculated using $\gamma = \Gamma / 2 + \Gamma^n$. We extracted the anharmonicity between the $\ket{0} \leftrightarrow \ket{1}$ and $\ket{1} \leftrightarrow \ket{2}$ transitions from two-tone spectroscopy~\cite{hoi2013microwave}. The anharmonicity approximately equals the charging energy $E_C$~\cite{Koch2007}. $\phi$ is the mismatch angle due to the impedance mismatch from the transmission lines.
\label{tab:1}}
\end{table*}

In \figpanel{fig:2}{b}, the transmission coefficient is depicted in the IQ plane for weak probe power $P_p$, where the Rabi frequency $\Omega$ is significantly smaller than the decoherence rate $\gamma$, resulting in a circular pattern. The experimental data points are indicated by red markers, while the solid curve represents the theoretical circle fit~\cite{Probst2015, lu2021characterizing}, from which we extract the parameters $\omega$, $\Gamma$, and $\gamma$, all summarized in \tabref{tab:1}. Note that we have accounted for impedance mismatch $\phi$ in both the theoretical curve and the experimental data in \figpanels{fig:2}{b}{c}.

Additionally, \figpanel{fig:2}{c} displays the magnitude of the transmission coefficient, $\abs{t}$, as a function of $P_p$ at the resonance frequency of the qubit. For low probe power, the incident coherent field undergoes extinction by up to \unit[94.7]{\%} in amplitude. With increasing probe power, a strong non-linearity emerges, saturating the qubit at high probe powers and leading to a unity transmission coefficient~\cite{Hoi2011}. The experimental data points are marked with red markers, and the theoretical fit is shown with a solid curve. All extracted parameters are summarized in \tabref{tab:1}.


\section{Calibration measurements in the time domain}
\label{sec:3}

To calibrate the gain and the attenuation of the input-output lines, we employ a time-domain setup as illustrated in \figref{fig:1} and perform single-tone calibration measurements. We bias our qubit at a resonance frequency of $\omega/ 2 \pi = \unit[4.1108]{GHz}$ and apply a pulse with a duration of $\unit[2]{\mu s}$. This duration is chosen to ensure that the qubit reaches a steady state, enabling us to obtain the average on-resonance signal in the steady state, ${\mleft\langle V_{t,\text{on}} \mright\rangle}$, across various probe powers. Similarly, we conduct experiments with the qubit significantly far detuned to measure the average off-resonance signal ${\mleft\langle V_{t,\text{off}} \mright\rangle}$. The magnitude of the transmission coefficient $\abs{t}$ is calculated as
\begin{equation}
\abs{t} = \abs{\frac{\mleft\langle V_{t,\text{on}} \mright\rangle}{\mleft\langle V_{t,\text{off}} \mright\rangle}} .
\label{eq:t}
\end{equation}
%

\begin{figure}
\includegraphics[width=\linewidth]{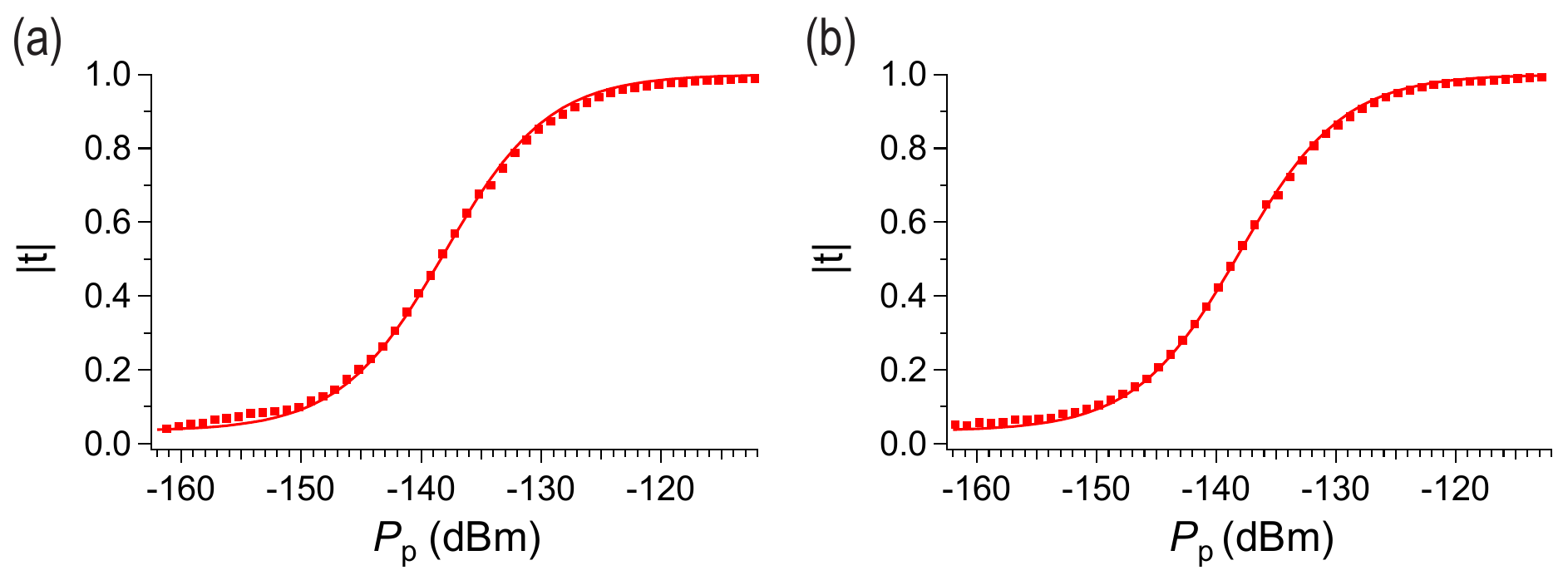}
\caption{The magnitude of the transmission coefficient, $\abs{t}$, as a function of probe power $P_p$ (converted from time domain) when the probe is resonant with the qubit. The figure displays both measured data points (markers) and theoretical fits (solid curves).
(a) Measurements taken on digitizer channel $A$, with the input port located on the left side of the AWG, indicated by $V_1$ (green arrow) in \figref{fig:1}.
(b) Measurements taken on digitizer channel $B$, with the input port located on the right side of the AWG, indicated by $V_2$ (blue arrow) in \figref{fig:1}.
\label{fig:3}}
\end{figure}

\begin{table}
\begin{tabular}{|c|c|c|c|c|c|}
\hline
\multicolumn{2}{|c|}{Frequency domain} & \multicolumn{4}{c|}{Time domain} \\
\hline
$A_{\rm freq}$ & $G_{\rm freq}$ & $A_{\rm ChA}$ & $G_{\rm ChA}$ & $A_{\rm ChB}$ & $G_{\rm ChB}$\\
\hline
$\unit[-125.05]{dB}$ & \unit[69.13]{dB} & $\unit[-132.06]{dB}$ & $\unit[89.77]{dB}$ & $\unit[-132.82]{dB}$ & $\unit[92.22]{dB}$ \\
\hline
\end{tabular}
\caption{Extracted gain $G$ and attenuation $A$ from frequency- and time-domain measurements~\cite{Cheng}. In the time-domain results, subscripts with ChA refer to calibration measurements taken on digitizer channel A with the input port for $V_1$, as shown in \figpanel{fig:3}{a}. Subscripts with ChB refer to measurements taken on digitizer channel B with the input port for $V_2$, as shown in \figpanel{fig:3}{b}.
}
\label{tab:2}
\end{table}

Figure~\figpanelNoPrefix{fig:3}{a} presents measurement results across different probe powers, where the input port is connected to the left side of the AWG, as indicated by the green arrow, and the transmitted output is captured on digitizer channel $A$ in \figref{fig:1}. Conversely, \figpanel{fig:3}{b} features the input connected to the right side of the AWG, as indicated by the blue arrow in \figref{fig:1}, with the transmitted output captured on digitizer channel $B$. The experimental data points are marked in red, while the solid curves represent theoretical fits using Eqs.~(4) and (5) in the main text. Based on the method described in Ref.~\cite{Cheng}, we can determine the gain and attenuation in time-domain measurements. For frequency-domain measurements, using the same fitting method as in \figpanel{fig:2}{c}, we can also determine the gain and attenuation. All extracted gain and attenuation are summarized in \tabref{tab:2}.


\section{Calculation of the scattering matrix}
\label{sec:4}

In this section, we discuss the calculations for the elements of the scattering matrix. Initially, we detune the qubit far away, resulting in full transmission ($t = 1$) for diagonal terms and no reflection ($r = 0$) for off-diagonal terms. The scattering matrix can then be expressed as
\begin{equation}
\begin{pmatrix}
V_3 \\
V_4
\end{pmatrix}
=
\begin{pmatrix}
1 & 0 \\
0 & 1
\end{pmatrix}
\begin{pmatrix}
V_1 \\
0
\end{pmatrix} 
\end{equation}
when the only input is the $V_1$ pulse ($V_2 = 0$). Here, $V_3 = V_1$, as depicted in Fig.~1(a) of the main text. Similarly, we obtain $V_4 = V_2$ when $V_1 = 0$:
\begin{equation}
\begin{pmatrix}
V_3 \\
V_4
\end{pmatrix}
=
\begin{pmatrix}
1 & 0 \\
0 & 1
\end{pmatrix}
\begin{pmatrix}
0 \\
V_2
\end{pmatrix} .
\end{equation}

When the qubit interacts with the field, the scattering matrix includes $t_L$ and $t_R$ as diagonal terms, and $r_L$ and $r_R$ as off-diagonal terms. The subscripts ``R" and ``L" denote ``right" and ``left", respectively, referring to the direction relative to the sample. The scattering matrix becomes
\begin{equation}
\begin{pmatrix}
V_3 \\
V_4
\end{pmatrix}
=
\begin{pmatrix}
t_L & r_R \\
r_L & t_R
\end{pmatrix}
\begin{pmatrix}
V_1 \\
0
\end{pmatrix} .
\end{equation}
when only the $V_1$ pulse is used as input; we then obtain $ V_3 = t_L V_1 $ and $V_4 = r_L V_1 $. Conversely, with $ V_1 = 0 $ and $ V_2 = V_2 $, we find $V_3 = r_R V_2 $ and $ V_4 = t_R V_2 $:
\begin{equation}
\begin{pmatrix}
V_3 \\
V_4
\end{pmatrix}
=
\begin{pmatrix}
t_L & r_R \\
r_L & t_R
\end{pmatrix}
\begin{pmatrix}
0 \\
V_2
\end{pmatrix}
\end{equation}

Through these calculations, we derive the coefficients $r_R$, $t_R$, $r_L$, and $t_L$, representing the right reflection, right transmission, left reflection, and left transmission coefficients, respectively, within the scattering matrix. These parameters are summarized in Table~\uppercase\expandafter{\romannumeral2} in the main text for Fig.~3(a). These coefficients provide valuable insights for evaluating the interference effects between the input pulses. Utilizing these coefficients, we simulate the theoretical results, allowing us to compare and validate experimental observations with theoretical predictions.


\section{Beam-splitter switching in the time domain}
\label{sec:5}

\begin{figure}
\includegraphics[width=\linewidth]{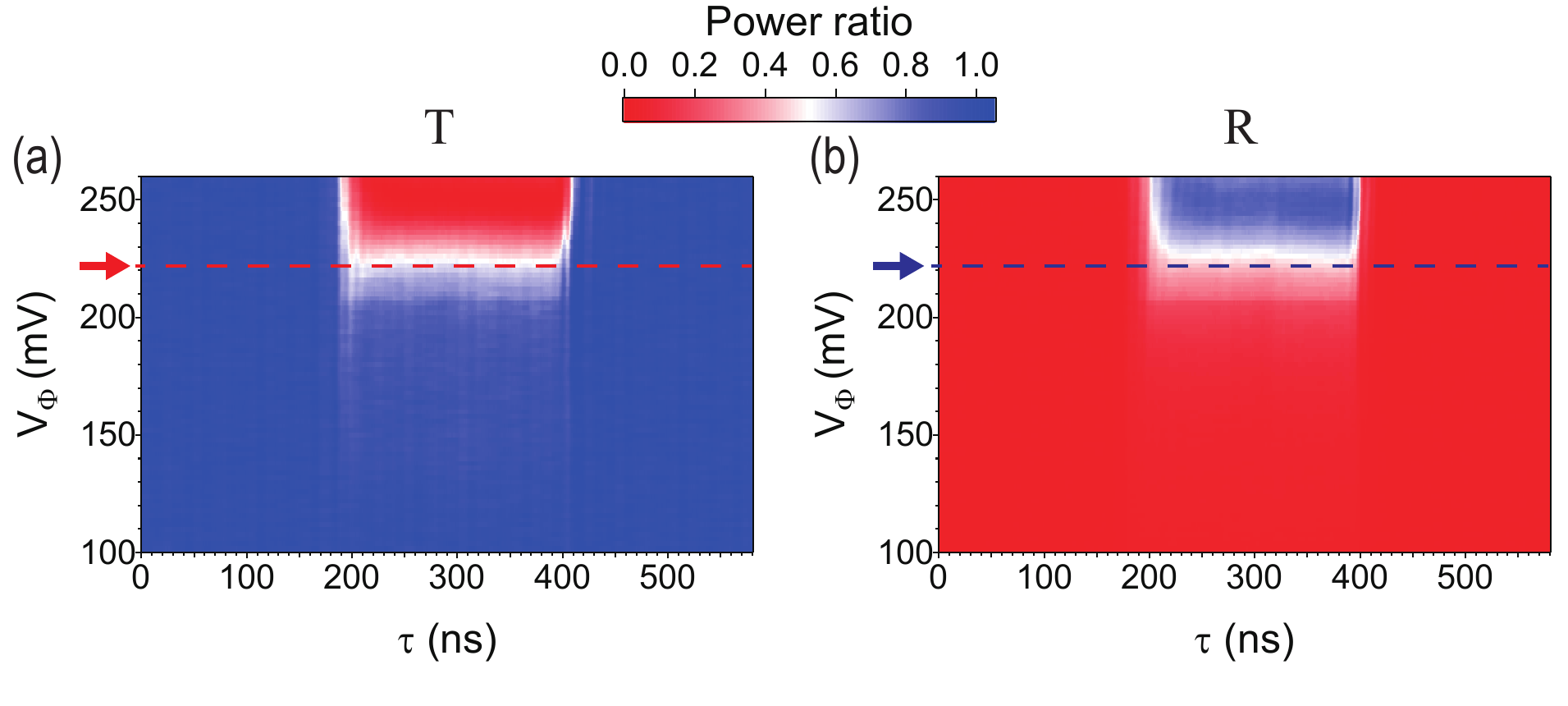}
\caption{Power ratio (transmittance and reflectance) for the output signals as a function of time and of the flux-biasing amplitude $V_\Phi$ at weak probe power ($\Omega \ll \gamma$).
(a) Transmittance $T$ measured using digitizer channel $A$. 
(b) Reflectance $R$ measured using digitizer channel $B$. A continuous wave is directed through one port of the open transmission line (indicated by a green arrow in \figref{fig:1}) with a probe frequency $\omega_p / 2 \pi = \unit[4.1108]{GHz}$ at a weak probe power of $\unit[-158.98]{dBm}$. Simultaneously, a Gaussian pulse is applied to the local flux of the qubit (shown by the brown arrow in \figref{fig:1}), with its voltage amplitude varying along the y axis, to adjust the qubit's resonance frequency, changing T and R between zero and unity (top color bar). The dashed line marks the line cut used in the main text as Fig.~2(b).
\label{fig:4}}
\end{figure}

In this section, we present the detailed results of Fig.~2(b) from the main text, shown in \figref{fig:4}. First, we biased the qubit at $\omega / 2\pi = \unit[4.2108]{GHz}$ (far detuned). Then, we varied the local flux by applying Gaussian pulses of varying amplitudes $V_\Phi$, as depicted on the y axis in \figref{fig:4}. We continuously probed the system using a continuous wave at a frequency of $\omega_p / 2\pi = \unit[4.1108]{GHz}$ with weak probe power. The pulse length is represented on the x axis.

In \figpanels{fig:4}{a}{b}, a noticeable change in the power ratio (indicated by a shift in color) occurs as we sweep the amplitude of the local flux, which adjusts the $\ket{0} \leftrightarrow \ket{1}$ transition frequency of the qubit. Near \unit[220]{mV} (indicated by a dashed line), we reach a 50:50 ratio (meaning that the system acts as a 50/50 beam splitter), similar to what is observed when the qubit is biased at $\omega / 2\pi = \unit[4.12]{GHz}$ in Fig.~2(a) of the main text (highlighted by a purple arrow). As we further increase the amplitude of the local flux, the power ratio shifts towards 0 and 1, corresponding to the qubit being on resonance at $\omega / 2\pi = \unit[4.1108]{GHz}$, with complementary values appearing in the lower region where the qubit is far detuned. This technique allows for switching the beam splitter in nanoseconds.


\section{Detailed data for the beam combiner around $T = 0.5$}
\label{sec:6}

\begin{figure}
\includegraphics[width=\linewidth]{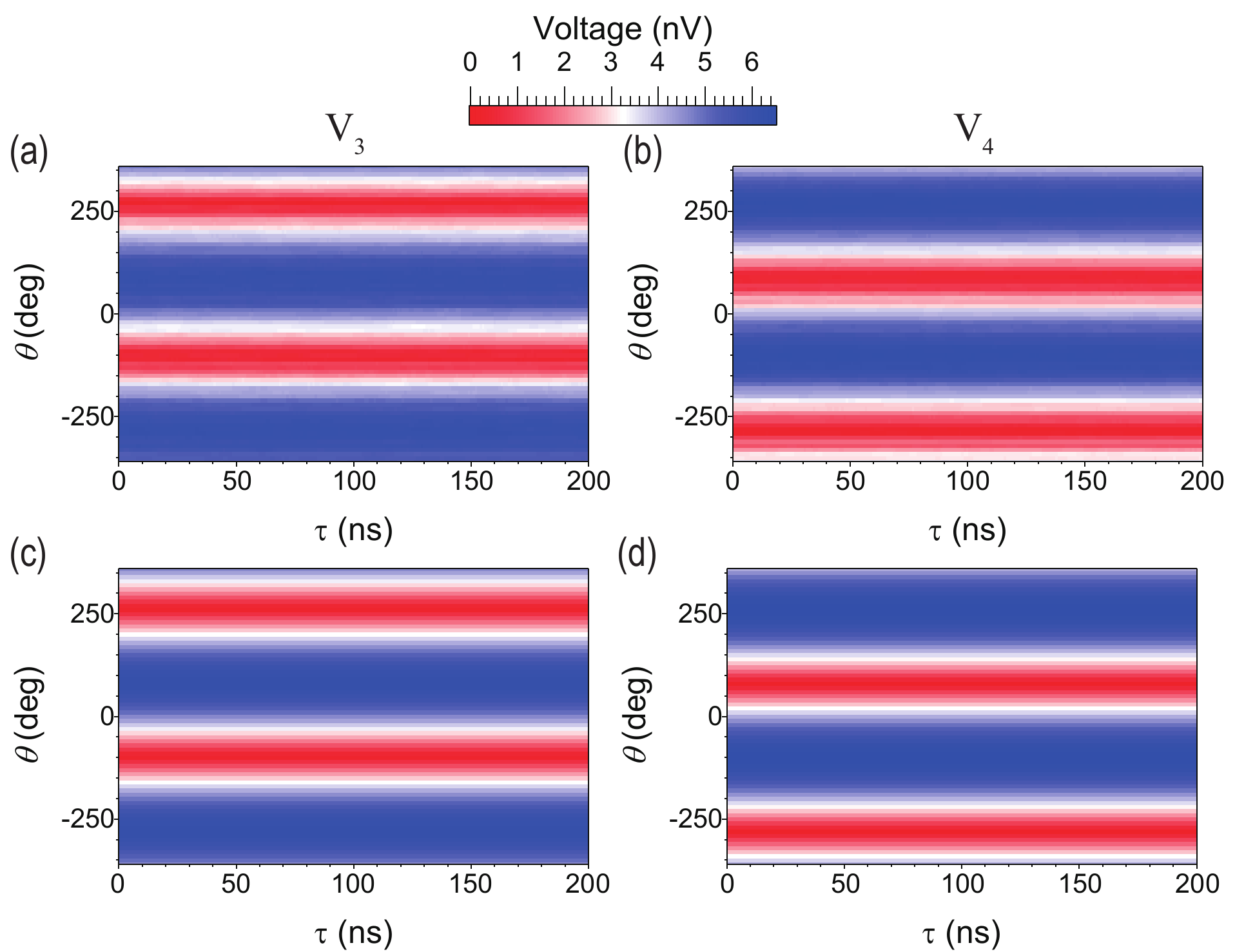}
\caption{Beam-combiner results around $T = 0.5$, as a function of time $\tau$ and the phase $\theta$ of the input $V_1$. 
(a) $V_3$ output signal measured using digitizer channel A.
(b) $V_4$ output signal measured using digitizer channel B. 
(c, d) Theoretical simulations corresponding to panels (a) and (b), respectively, based on Eqs.~(1)--(5) from the main text, using the parameters listed in Table~\uppercase\expandafter{\romannumeral2} of the main text.
\label{fig:5}}
\end{figure}

In this section, we present the detailed interference data of Fig.~3(a) in the main text. As mentioned, we set the qubit's resonance frequency to $\omega / 2\pi = \unit[4.12]{GHz}$ to achieve a transmittance and reflectance ratio of approximately 50:50. Two weak probes (labeled $V_1$ and $V_2$) at $\omega_p / 2\pi = \unit[4.1108]{GHz}$ are simultaneously applied to the atom with phase $\theta$ for $V_1$. The output signals $V_3$ ($V_4$) are then measured using digitizer channel A (B), as shown in \figpanels{fig:5}{a}{b}.

We extract and average the steady-state signal from the data corresponding to $\tau$ from 50 to \unit[150]{ns} for each $\theta$ case. This yields the blue (red) data points in Fig.~3(a) in the main text. The experimental data matches well with the simulation results in \figpanels{fig:5}{c}{d}.


\clearpage

\newpage

\bibliography{sup}